**Automating Freshman Course Placement and Registration: A Case Study**

Bharathwaj Vijayakumar[1], Samyukta Alapati[2], Sahana Varadaraju[3]

[1]Assistant Vice President, Office of Institutional Research & Analytics, Rowan University, Glassboro, NJ 08028 USA, vijayakumar@rowan.edu
[2]Director, Office of Institutional Research & Analytics, Rowan University, Glassboro, NJ 08028 USA, alapati@rowan.edu
[3]Senior Cloud Application Developer, Office of Institutional Research & Analytics, Rowan University, Glassboro, NJ 08028 USA, varadaraju@rowan.edu

**Abstract**

This implementation report explores Rowan University's effort to automate the process of freshman course placement and registration. Historically, Freshman Instructional Guides (FIGS) at Rowan was manually executed, requiring significant time from Testing Services, University Advising, and the Registrar's Office to evaluate placement needs and assign students to courses. Given the 57% surge in first-time degree-seeking student enrollment over a decade, the manual processes became increasingly unsustainable. In response, a cross-departmental team developed a comprehensive automated process to integrate data from Banner (Student Information System), Google Sheets maintained by Advising, and other sources. This automated process classifies students based on program groupings, determines primary and secondary course placements, checks for real-time availability and constraints in Banner, and completes course registration for freshmen in bulk. The resulting system processed over 3500 incoming students with over 350 hours in annual time savings, reduced the potential for human error, and enabled staff to shift focus from administrative work to strategic advising. This report outlines the implementation context, design architecture, technical integration, assessment methods, lessons learned, and practical implications for institutions with similar challenges.

## Introduction and Implementation Context

Rowan University, located in Glassboro, New Jersey, serves over 22,000 students across a diverse range of academic programs. With a 57% increase in first-time degree-seeking (FTDS) enrollment over the past decade, Rowan's traditional Freshman Instructional Guides (FIGS) process for onboarding new students became increasingly difficult to manage. This manual approach interpreting placement scores, reviewing transfer credits, and assigning foundational courses like math, chemistry, and English was carried out collaboratively by Testing Services, University Advising, and the Registrar's Office. Similar bottlenecks have been observed at other institutions facing enrollment growth without scalable registration systems (Crosta, 2013).

In response, Rowan University implemented a technology-driven solution to automate freshman placement and registration. Automated registration tools have helped institutions reduce delays and streamline scheduling, especially when student volume and placement complexity increase (Tilahun & Sekeroglu, 2020). This approach aligns with guided pathways models, which emphasize proactive, structured course enrollment as a driver of student momentum. Early course placement and sequence planning can also reduce credit loss and support degree completion (Bailey, Jaggars, & Jenkins, 2015).

The automation effort at Rowan was developed by a cross-functional team from Office of Institutional Research & Analytics (OIRA), Registrar, Business Intelligence, Software

Development, Testing Services, and Advising. The team's goal was to create a scalable, rules-based system that enhanced placement accuracy and aligned with the Rowan's broader mission to improve advising efficiency, expand course access, and support student success, an approach consistent with the principles outlined in Bailey et al. (2015) and (Tilahun & Sekeroglu, 2020)..

Similar automation initiatives have emerged across higher education as institutions confront the growing complexity of placement and registration. For example, Arizona State University and Georgia State University have implemented guided, data-driven onboarding models that integrate advising rules directly into registration workflows (Dadger et al., 2018).

The University of Central Florida has also adopted an automated degree audit and registration framework that uses predictive analytics to preassign courses based on student pathways (University of Central Florida, 2022). These initiatives reflect a national shift toward digital transformation and process automation in student services (Tyton Partners, 2022; EDUCAUSE Horizon Report, 2024). They emphasize scalability, transparency, and equity through technology-enabled advising (Nixon, 2021).

The approach outlined in this case study contributes to this growing body of practice by demonstrating how a university can use existing systems such as Banner and advising-managed datasets to achieve comparable efficiencies without reliance on external vendors or large enterprise systems.

## Background and Rationale

Before this automation effort, placement and registration were managed using static spreadsheets, manual comparisons of Banner data, and individual case reviews. Testing Services downloaded standardized test scores and transcript data, applying institutional placement rules through Excel-based workflows. Advisors then matched placement outcomes to appropriate course levels, carefully considering co-requisites and program-specific restrictions, student by student.

At its peak, this process consumed over 350 combined staff hours annually and was vulnerable to delays, inconsistencies, and human error. With freshman enrollment surpassing 3500 students in fall 2024, this kind of manual effort became very time consuming and error prone. Errors such as incorrect placement or unnecessary testing could impact student satisfaction and progression.

Recognizing the growth and risk involved in error prone manual assignments, Rowan sought a centralized, scalable solution aligned with its goals for student success, digital infrastructure, and institutional agility. The project brought together multiple functional and technical units to develop a sustainable automation framework that would modernize and streamline freshman registration.

## Implementation Approach

The implementation followed a structured yet agile development framework as shown in Fig 1 and Fig 2. Initial planning began in Spring 2021, with a pilot rollout in Fall 2021 and iterative improvements through 2025.

Advising and Testing Services defined the placement logic, including exemption thresholds for reading, writing, math, and chemistry. For example, a student could bypass the reading placement test with a SAT Reading Subscore of 27+, ACT Reading of 21+, or 30+ transfer credits. Math placement rules varied by major; STEM programs required higher thresholds (e.g., SAT Math 620+), while other programs used alternative criteria. These rules, including Accuplacer cutoffs and transcript-based overrides, were embedded into a rules engine to ensure equitable and consistent placement.

This process categorized programs into groups A–E based on an advising-maintained Google Sheet, each with unique course assignment rules. Placement logic determined whether a student required testing, what course level was appropriate, and which primary and secondary course options should be assigned. Once eligibility and placement were determined, the registration module validated availability, checked for scheduling conflicts, and ensured compliance with cohort or major-based restrictions before enrolling students in Banner. Any exceptions such as full sections or ineligible students were flagged for manual review. Weekly meetings were held to test logic, resolve flagged cases, and iterate improvements. All logic rules and placement mappings were version-controlled for transparency and long-term maintainability.

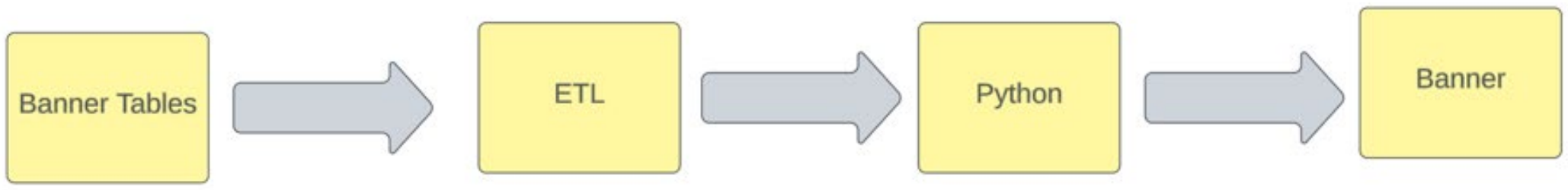

Fig 1. Overall Flow of FIGS Process

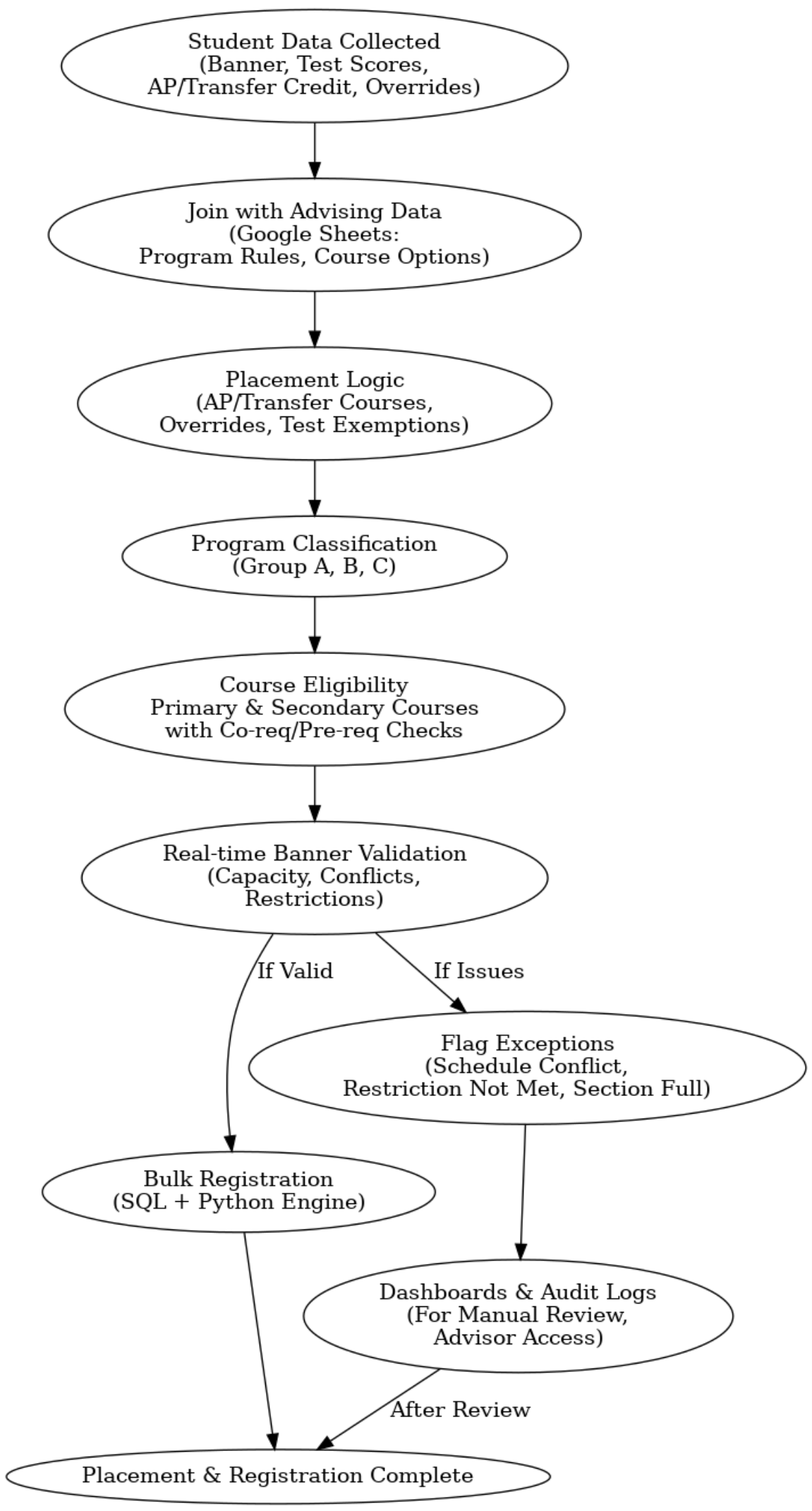

Fig 2. Figs Process Details

## Technical Implementation Details

The system's architecture comprised multiple modular Extract, Transform & Load (ETL) jobs orchestrated to process student placement and registration in stages. At the core of the system was an ETL pipeline built using Veera Construct, Rowan's enterprise data integration platform. As illustrated in *Appendix 1*, the pipeline extracted student data from Banner including test scores, transfer credits, program codes, and demographic information and joined it with external sources like advising-maintained Google Sheets. The logic engine, built in Veera Construct, applied over 30 program and score-specific placement rules. It assigned students to program groups (A–E), evaluated exemption eligibility, and selected course groupings based on predefined advising templates. This included logic for dual placement, developmental coursework, and exceptions.

The final stage in the pipeline passed course assignments to a bulk registration engine. This component dynamically queried Banner to verify section availability, identify restrictions such as cohort attributes or major restrictions, and confirm there were no scheduling conflicts. For example, if a course section had already reached capacity, the student would be flagged for manual registration review. Similarly, if a course was restricted to a specific student cohort such as honors students or first-year learning community members or limited to students in a specific major, the engine would evaluate eligibility before proceeding. Time conflict checks ensured that assigned sections did not overlap with other registered courses. If all placement and eligibility conditions were satisfied, the student was automatically enrolled in the designated courses within Banner. Any records with conflicts or unmet conditions were flagged with a descriptive error code (e.g., "Section Full," "Major Restriction Not Met," or "Schedule Conflict") for advisor follow-up and resolution. The system supports incremental data refreshes and runs nightly, ensuring updates from Banner or advising documents are incorporated quickly. Built-in audit logs track rule application, registration status, and exception categories, ensuring transparency and maintainability.

## Evaluation Methods and Metrics

To ensure the system performed as expected, a series of detailed test cases were developed prior to rollout. These cases covered a wide variety of student profiles, including students in specific programs such as EOF, students requiring COMP (composition) placements, and those needing specialized 101 sections. The complexity of certain programs and placement pathways demanded precise evaluation to confirm that rules and logic were correctly applied.

The Business Intelligence and Advising teams validated the system by comparing expected outcomes against the automated results across these scenarios. As shown in *Appendix 2 and 3*, Dashboards and audit reports were developed to display individual student placements, registration assignments, and test requirements. Advisors and Testing Services teams then reviewed these dashboards to confirm placement accuracy and provide feedback.

Additionally, the Registrar's Office contributed to testing by verifying whether the automated enrollments respected course restrictions, section attributes, and capacity constraints. Any anomalies were flagged, reviewed collaboratively, and addressed through rule refinement.

This iterative validation approach combining targeted test cases, visual dashboards, and stakeholder review ensured the system met operational and academic expectations. Updates and corrections were incorporated into each registration cycle, and findings were reported back to the core development team for continuous improvement.

## Results and Outcomes

The most immediate benefit of the implementation was a substantial reduction in manual workload across multiple departments. Testing Services reported saving over 150 hours per year, University Advising reclaimed approximately 100 hours, and the Registrar's Office saved around 100 hours of time that was previously devoted to reviewing placement data and matching students to appropriate courses. These savings enabled staff to redirect their efforts toward proactive advising, academic coaching, and strategic planning.

Beyond operational efficiency, the system demonstrated a high degree of accuracy in placement determinations, significantly reducing manual errors. Advisors expressed greater confidence in the results, and the transparent, rule-based decision logic allowed them to understand and explain placement outcomes with clarity. This enhanced trust in the system and improved the overall advising experience.

A key academic benefit of accurate placement is its role in supporting timely degree progression. When freshmen are placed directly into the correct foundational courses, they are more likely to complete key prerequisites early. This enables them to access higher-level major courses in the appropriate semesters, reducing the risk of delays in program completion. Incorrect placements either too advanced or too remedial can lead to wasted semesters, unnecessary credits, or gaps in academic momentum.

The automation also improved curriculum planning capabilities for academic departments. By aggregating and applying placement logic earlier in the enrollment cycle, departments could better forecast course demand, especially for high-enrollment or sequential courses. Preliminary registration rosters became available weeks in advance, giving academic schedulers more time to open additional sections, identify bottlenecks, and reduce last-minute changes that disrupt student schedules.

This system not only streamlined internal processes but also created conditions for stronger student outcomes. Accurate, early placement into the right courses puts students on a clearer academic path from the outset, helping them stay on track, avoid course sequencing delays, and ultimately graduate on time.

## Challenges and Solutions

One of the key challenges during implementation was handling inconsistencies in student data, particularly when test scores conflicted with transfer credit or when students change majors mid-process. The team addressed this by introducing nightly refresh logic and creating exception alerts.

Another challenge was building trust in the automated decisions. To address this, the project team hosted walkthroughs, opened the logic to advisor review, and provided override mechanisms. Over time, stakeholders expressed increased confidence in the system's outputs.

There were also technical hurdles related to integration with Banner. These included aligning registration logic with Banner's enforcement of section restrictions and student eligibility flags. To overcome this, the team developed custom queries that adhered to institutional data definitions and ensured the automation respected Banner's structure for course prerequisites, cohort rules, and attribute-based permissions. Ongoing collaboration with the Registrar's Office helped verify that the registration logic honored existing workflows and that any exceptions were addressed early in the process.

## Lessons Learned and Best Practices

Several key lessons emerged. First, stakeholder engagement throughout the process ensured higher adoption and better alignment with institutional needs. Second, transparent rule documentation facilitated trust and ongoing maintenance. Third, designing the system to be flexible and easily updated by non-technical users improved sustainability. Training new staff using the dashboards proved to be more efficient than onboarding them to spreadsheet-based methods. The implementation reaffirmed the value of automating repeatable, rule-driven processes and demonstrated that institutional data systems can be leveraged creatively to support student success. Fourth, cross-functional governance is essential. Establishing a standing “Placement Automation Team” with representation from Advising, Testing Services, and IT ensured ongoing rule accuracy and quick resolution of exceptions. Another important lesson involved managing unanticipated data issues that emerged during implementation. Early iterations revealed that even small inconsistencies such as missing or late updates to cohort codes, major-based section restrictions, or attribute tags in Banner could prevent eligible students from being placed or registered automatically. Ensuring that these data elements were entered accurately and on schedule proved critical to the system’s success. The project team learned to coordinate closely with the Registrar’s scheduling timeline and to include validation checks that flag missing or outdated restriction codes before each automation run. Establishing clear data governance practices and accountability for these updates was essential for sustaining reliability and scalability.

## Sustainability and Future Development

Recent developments in higher education technology highlight the growing role of artificial intelligence and process automation in supporting student onboarding and advising. Institutions are increasingly using AI-driven logic to interpret placement criteria, generate course recommendations, and automate registration decisions (Rawatlal & Dhunpath, 2023; Thottoli et al., 2024; Soomro et al., 2025)). These trends demonstrate how the next generation of placement systems will combine data governance, analytics, and intelligent automation to enhance equity, transparency, and scalability across academic operations.

The current system at Rowan is sustained through shared ownership across OIRA, Business Intelligence, Software Development, University Advising, and Testing Services. Placement rules are reviewed annually and adjusted to reflect updated policies or curricular changes. The architecture remains flexible and supports scalable updates without requiring complete redevelopment.

Looking forward, the next major evolution involves integrating generative AI to simplify and scale rule deployment. Instead of manually configuring placement logic in ETL workflows, users will be able to provide a structured document containing rules such as exemption criteria or course eligibility conditions which the AI will interpret and apply dynamically. This will reduce maintenance overhead, accelerate onboarding for new programs, and provide greater transparency.

Additional enhancements under consideration include real-time scheduling conflict detection, integration with academic advising tools, and a student-facing chatbot capable of explaining placement outcomes and offering tailored support resources. These tools aim to further improve the student experience and align placement with evolving institutional strategies. Annual evaluations will continue to guide improvements and ensure that the system remains responsive to academic needs and stakeholder feedback

## Conclusion

This paper aligns with the Data Strategy and Institutional Intelligence thematic area by demonstrating how this Rowan transformed a manual, error-prone freshman course placement and registration process into an automated, data-driven system. The automation of this Rowan’s FIGS process represents a successful transformation of a high-volume, labor-intensive operation into a scalable, transparent, and accurate system. The result is a 350-hour reduction in staff time annually, improved placement accuracy, and enhanced capacity for student-centered advising. To quantify time savings, each participating office documented the average duration of key tasks during the 2020 cycle, including placement rule application, manual course matching, and registration verification. These baselines were compared against the automated process using the same number of students in 2020. The cumulative difference across Testing Services, Advising, and the Registrar's Office yielded an estimated 350 hours of annual time savings. This calculation was validated through staff time logs and system-generated processing metrics.

By embedding automation into a previously manual process, Rowan University was able to both improve operational efficiency and reinforce its commitment to student success. The collaboration across institutional units and the adaptability of the technical framework underscore the importance of investing in sustainable, data-driven process redesign. The initiative serves as a replicable model for institutions seeking to modernize registration and placement processes using their existing student information systems.

**Acknowledgments**

The authors would like to acknowledge the staff at Rowan University, including Dr. Rory McElwee, *Vice President for Student Affairs*; Jackie Ring, Vice Chancellor, Chief institutional Research OfficerAmy Ruymann, *Assistant Vice President for Academic Support, Advising, and Analytics*; Christopher Fields, *Assistant Registrar of Enrollment Reporting & Compliance*; Linda Drexel, *University Registrar*; Christopher Taylor, *Assistant Director for Student Success Technology and Data Analytics*; Isaac Strozyk, *Business Intelligence Analyst*; and Himadriben Patel, *Senior Business Application Developer*, for their contributions and collaboration in the development and implementation of the bulk registration process.

# References

Bailey, T. R., Jaggars, S. S., & Jenkins, D. (2015). *Redesigning America's community colleges: A clearer path to student success*. Harvard University Press.

Crosta, P. M. (2013). Intensity and attachment: How the chaotic enrollment patterns of community college students relate to educational outcomes. *Community College Review*, 41(2), 118–142. https://doi.org/10.1177/0091552112471556

Tilahun, L. A., & Sekeroglu, B. (2020). *An intelligent and personalized course advising model for higher educational institutes. SN Applied Sciences, 2*, Article 1635. https://doi.org/10.1007/s42452-020-03440-4

University of Central Florida. (2022). *2022 accountability plan.* Board of Governors, State University System of Florida. https://www.flbog.edu/wp-content/uploads/2022/07/UCF_2022_Accountability_Plan_BOG_Approved.pdf

EDUCAUSE. (2024). *Horizon Report: Teaching and Learning Edition.* EDUCAUSE. https://library.educause.edu/resources/2024

Nixon, C. T. (2021, April 13). Effective use of technologies in student advising: Is there a Yellow Brick Road? *EDUCAUSE Review*. https://er.educause.edu/articles/2021/4/effective-use-of-technologies-in-student-advising-is-there-a-yellow-brick-road

Tyton Partners. (2022). Time for Class 2022: The State of Digital Learning and Courseware Adoption. Tyton Partners Research Brief. https://tytonpartners.com

Dadgar, M., Fischerhall, C., Collins, L., & Schaefer, K. (2018). *Guided pathways as a framework for integrating student success efforts: A case study of three California community colleges*. Oakland, CA: Career Ladders Project.

Rawatlal, R., & Dhunpath, R. (2023). *Advising at scale: Automated guidance of the role players influencing student success.* The AIR Professional File, Fall 2023(163). https://doi.org/10.34315/apf1632023

Thottoli, M. M., Alruqaishi, B. H., & Soosaimanickam, A. (2024). *Robo academic advisor: Can chatbots and artificial intelligence replace human interaction?* Contemporary Educational Technology, 16(1), ep485. https://doi.org/10.30935/cedtech/13948

Soomro, A. A., Khan, M. H., Umar, M., Khan, S., & Ali, O. (2025). *AI-driven academic advising in higher education: Leveraging intelligent systems to personalize student support, improve retention, and optimize career pathways.* The Critical Review of Social Sciences Studies, 3(2), 229–248. https://doi.org/10.59075/vy3v7k17

## Appendices

### Appendix 1: Veera Contruct Job for Course Assignments

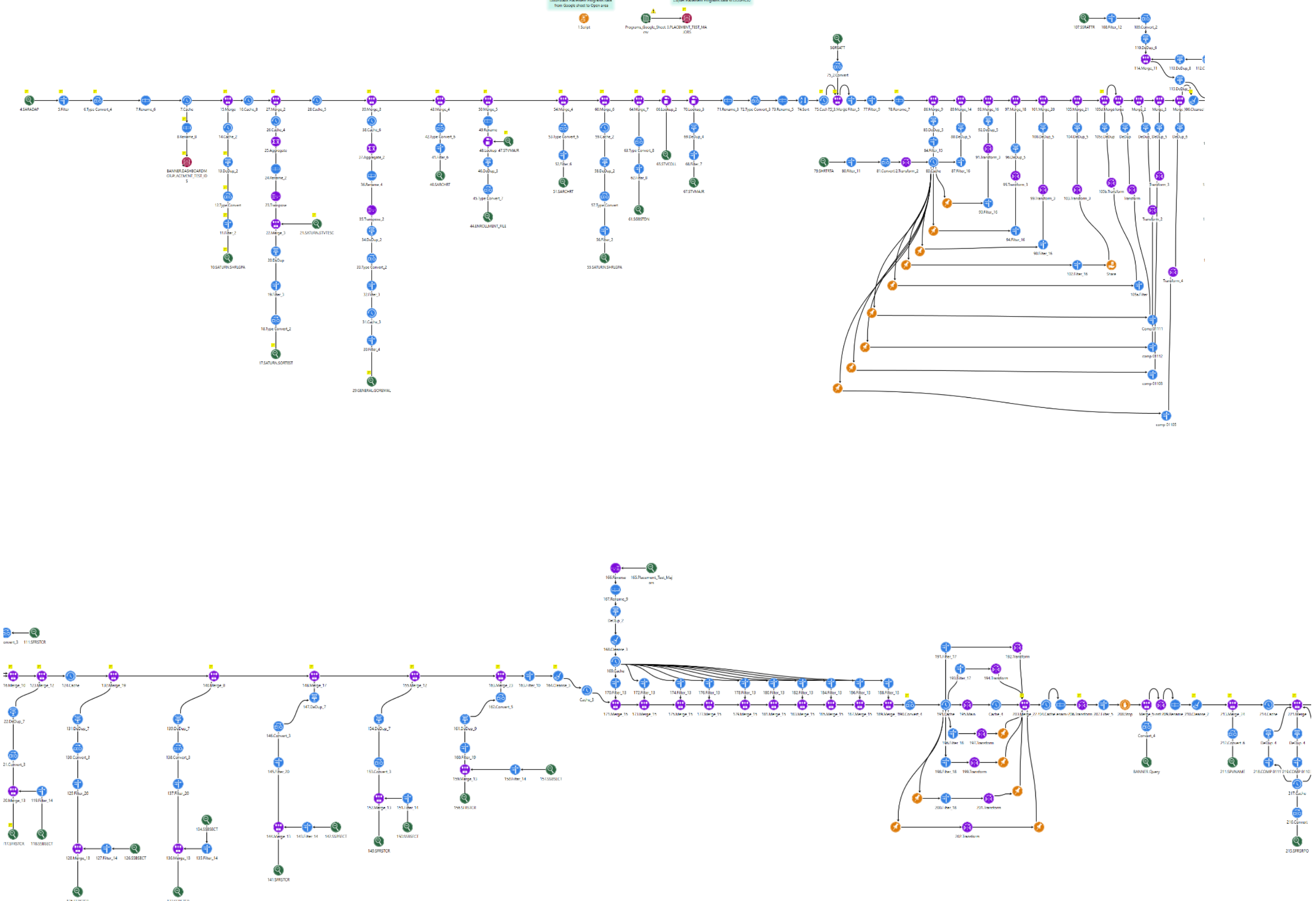

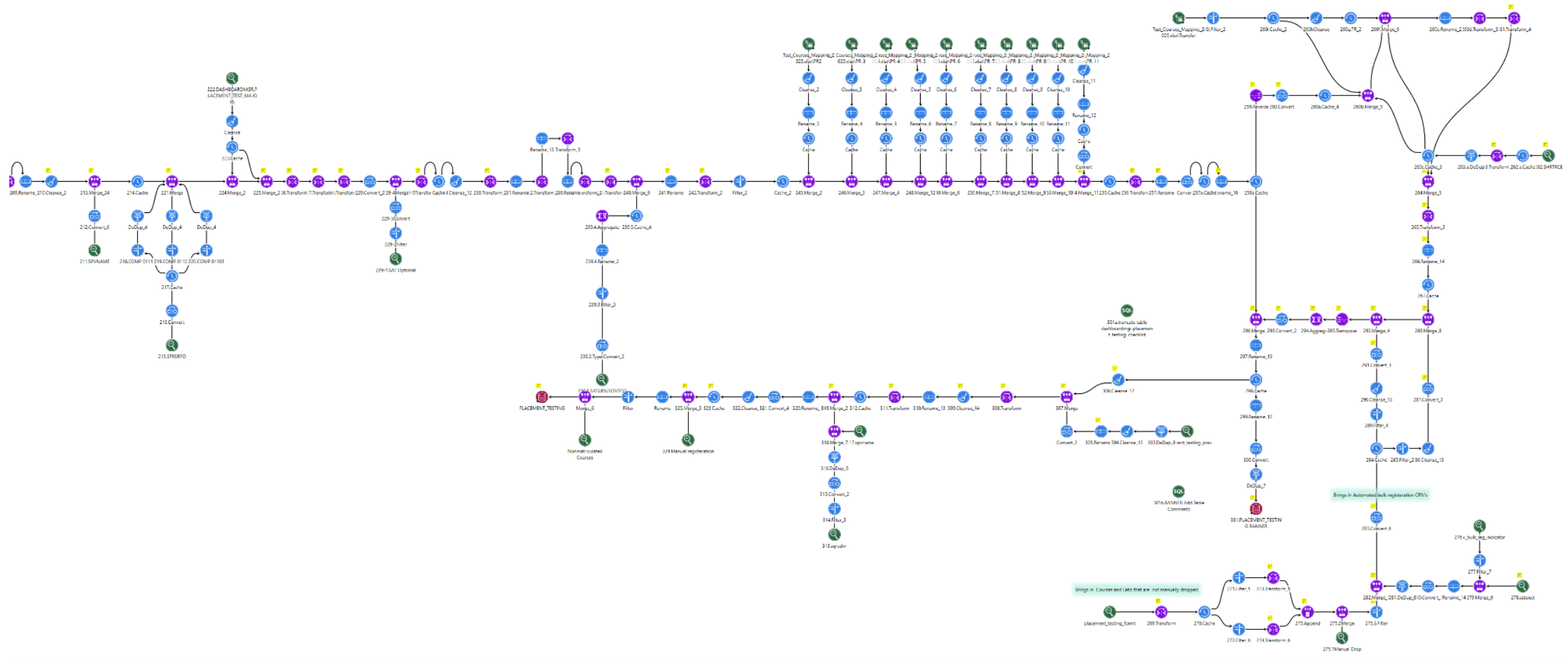

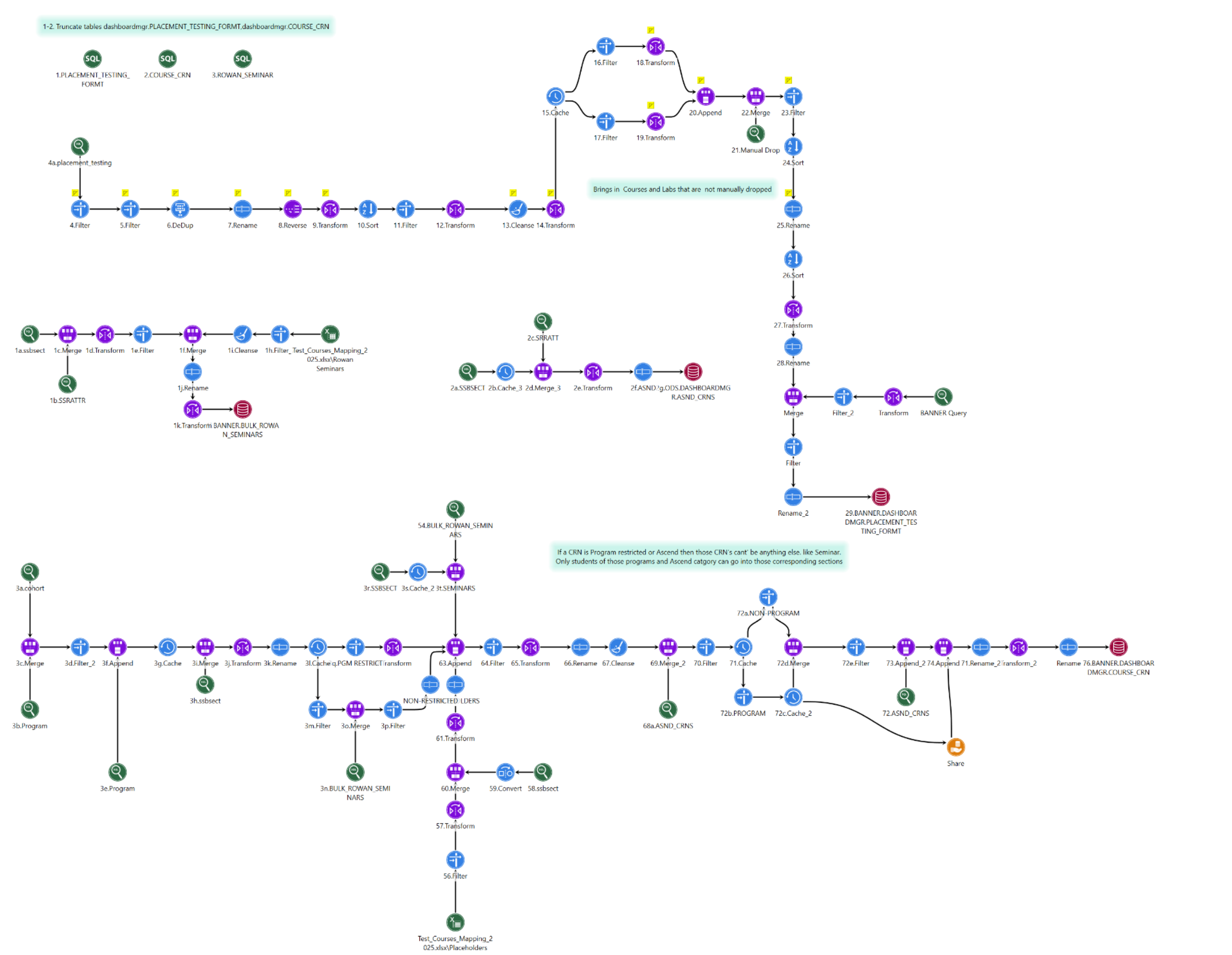

1-2. Truncate tables dashboardmgr.PLACEMENT_TESTING_FORMT.dashboardmgr.COURSE_CRN
1.PLACEMENT_TESTING_FORMT
2.COURSE_CRN
3.ROWAN_SEMINAR
4a.placement_testing
4.Filter
5.Filter
6.DeDup
7.Rename
8.Reverse
9.Transform
10.Sort
11.Filter
12.Transform
13.Cleanse
14.Transform
15.Cache
16.Filter
18.Transform
17.Filter
19.Transform
20.Append
22.Merge
23.Filter
21.Manual Drop
24.Sort
Brings in Courses and Labs that are not manually dropped
25.Rename
26.Sort
27.Transform
28.Rename
Merge
Filter_2
Transform
BANNER Query
Filter
Rename_2
29.BANNER.DASHBOARDMGR.PLACEMENT_TESTING_FORMT
1a.ssbsect
1c.Merge
1d.Transform
1e.Filter
1f.Merge
1i.Cleanse
1h.Filter_Test_Courses_Mapping_2025.xlsx\Rowan Seminars
1b.SSRATTR
1j.Rename
1k.Transform BANNER.BULK_ROWAN_SEMINARS
2c.SSRATT
2a.SSBSECT
2b.Cache_3
2d.Merge_3
2e.Transform
2f.ASND.fg.ODS.DASHBOARDMGR.ASND_CRNS
54.BULK_ROWAN_SEMINARS
If a CRN is Program restricted or Ascend then those CRN's cant' be anything else. like Seminar. Only students of those programs and Ascend catgory can go into those corresponding sections
3r.SSBSECT
3s.Cache_2
3t.SEMINARS
3a.cohort
3c.Merge
3d.Filter_2
3f.Append
3g.Cache
3i.Merge
3j.Transform
3k.Rename
3l.Cache
q.PGM RESTRICT
ransform
63.Append
64.Filter
65.Transform
66.Rename
67.Cleanse
69.Merge_2
70.Filter
71.Cache
72a.NON-PROGRAM
72d.Merge
72e.Filter
73.Append_2
74.Append
71.Rename_2
Transform_2
Rename
76.BANNER.DASHBOARDMGR.COURSE_CRN
3b.Program
3h.ssbsect
NON-RESTRICTED-LDERS
3m.Filter
3o.Merge
3p.Filter
61.Transform
68a.ASND_CRNS
72b.PROGRAM
72c.Cache_2
72.ASND_CRNS
Share
3e.Program
3n.BULK_ROWAN_SEMINARS
60.Merge
59.Convert
58.ssbsect
57.Transform
56.Filter
Test_Courses_Mapping_2025.xlsx\Placeholders

Appendix 2 : Placement Testing Dashboard

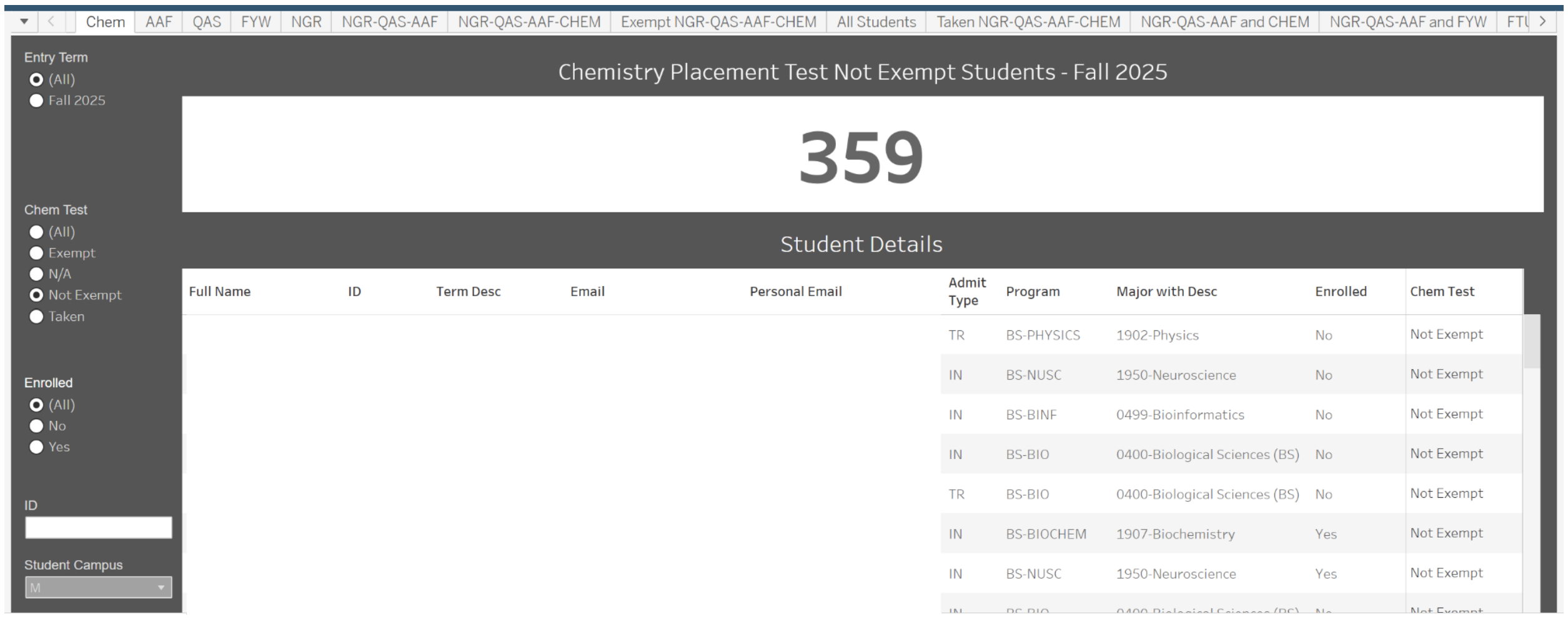

| Full Name | ID | Term Desc | Email | Personal Email | Admit Type | Program | Major with Desc | Enrolled | Chem Test |
|---|---|---|---|---|---|---|---|---|---|
| | | | | | TR | BS-PHYSICS | 1902-Physics | No | Not Exempt |
| | | | | | IN | BS-NUSC | 1950-Neuroscience | No | Not Exempt |
| | | | | | IN | BS-BINF | 0499-Bioinformatics | No | Not Exempt |
| | | | | | IN | BS-BIO | 0400-Biological Sciences (BS) | No | Not Exempt |
| | | | | | TR | BS-BIO | 0400-Biological Sciences (BS) | No | Not Exempt |
| | | | | | IN | BS-BIOCHEM | 1907-Biochemistry | Yes | Not Exempt |
| | | | | | IN | BS-NUSC | 1950-Neuroscience | Yes | Not Exempt |
| | | | | | IN | BS-BIO | 0400-Biological Sciences (BS) | No | Not Exempt |

Appendix 3: Bulk Registration Report

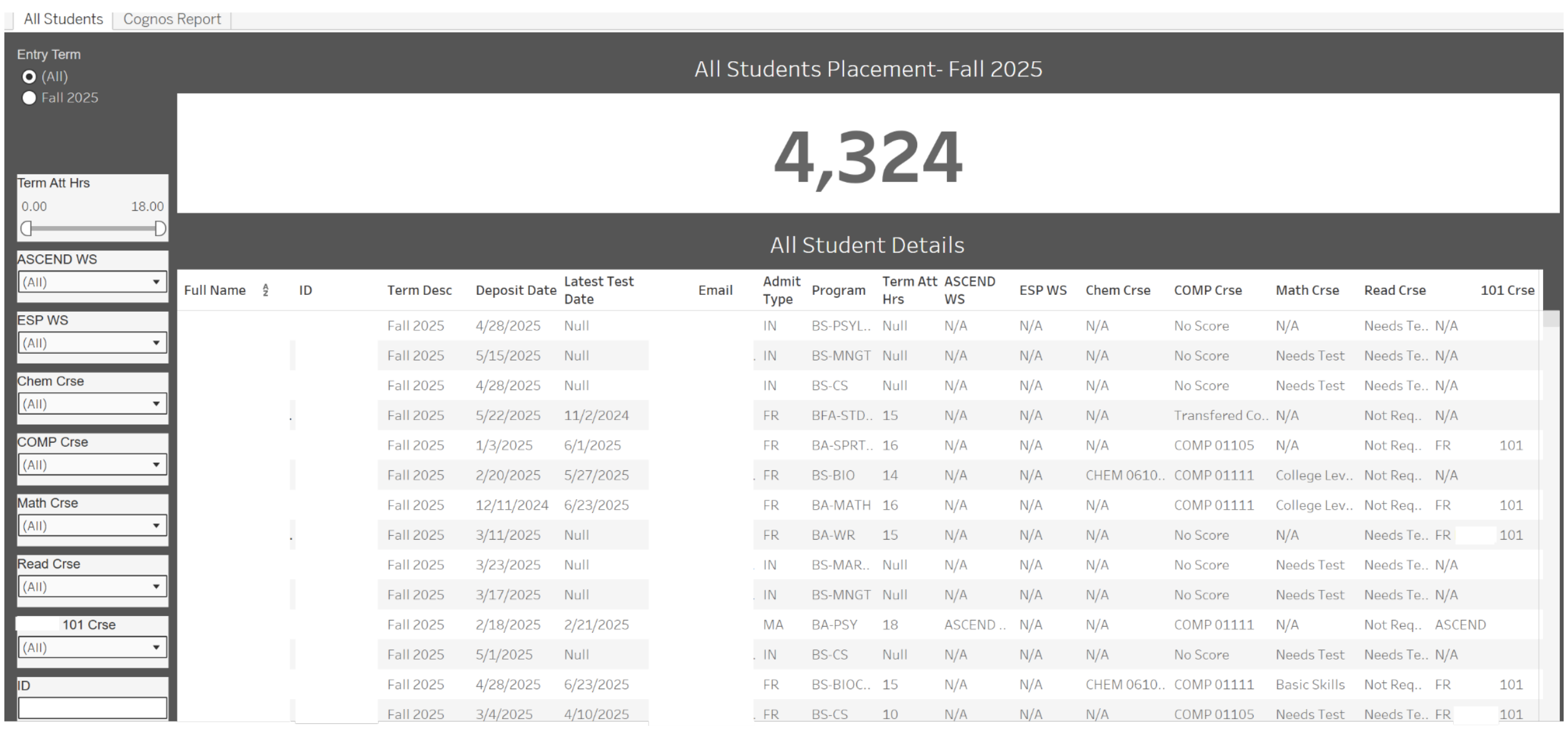

| Full Name | ID | Term Desc | Deposit Date | Latest Test Date | Email | Admit Type | Program | Term Att Hrs | ASCEND WS | ESP WS | Chem Crse | COMP Crse | Math Crse | Read Crse | | 101 Crse |
|---|---|---|---|---|---|---|---|---|---|---|---|---|---|---|---|---|
| | | Fall 2025 | 4/28/2025 | Null | | IN | BS-PSYL.. | Null | N/A | N/A | N/A | No Score | N/A | Needs Te.. | N/A | |
| | | Fall 2025 | 5/15/2025 | Null | | IN | BS-MNGT | Null | N/A | N/A | N/A | No Score | Needs Test | Needs Te.. | N/A | |
| | | Fall 2025 | 4/28/2025 | Null | | IN | BS-CS | Null | N/A | N/A | N/A | No Score | Needs Test | Needs Te.. | N/A | |
| | | Fall 2025 | 5/22/2025 | 11/2/2024 | | FR | BFA-STD.. | 15 | N/A | N/A | N/A | Transfered Co.. | N/A | Not Req.. | N/A | |
| | | Fall 2025 | 1/3/2025 | 6/1/2025 | | FR | BA-SPRT.. | 16 | N/A | N/A | N/A | COMP 01105 | N/A | Not Req.. | FR | 101 |
| | | Fall 2025 | 2/20/2025 | 5/27/2025 | | FR | BS-BIO | 14 | N/A | N/A | CHEM 0610.. | COMP 01111 | College Lev.. | Not Req.. | N/A | |
| | | Fall 2025 | 12/11/2024 | 6/23/2025 | | FR | BA-MATH | 16 | N/A | N/A | N/A | COMP 01111 | College Lev.. | Not Req.. | FR | 101 |
| | | Fall 2025 | 3/11/2025 | Null | | FR | BA-WR | 15 | N/A | N/A | N/A | No Score | N/A | Needs Te.. | FR | 101 |
| | | Fall 2025 | 3/23/2025 | Null | | IN | BS-MAR.. | Null | N/A | N/A | N/A | No Score | Needs Test | Needs Te.. | N/A | |
| | | Fall 2025 | 3/17/2025 | Null | | IN | BS-MNGT | Null | N/A | N/A | N/A | No Score | Needs Test | Needs Te.. | N/A | |
| | | Fall 2025 | 2/18/2025 | 2/21/2025 | | MA | BA-PSY | 18 | ASCEND .. | N/A | N/A | COMP 01111 | N/A | Not Req.. | ASCEND | |
| | | Fall 2025 | 5/1/2025 | Null | | IN | BS-CS | Null | N/A | N/A | N/A | No Score | Needs Test | Needs Te.. | N/A | |
| | | Fall 2025 | 4/28/2025 | 6/23/2025 | | FR | BS-BIOC.. | 15 | N/A | N/A | CHEM 0610.. | COMP 01111 | Basic Skills | Not Req.. | FR | 101 |
| | | Fall 2025 | 3/4/2025 | 4/10/2025 | | FR | BS-CS | 10 | N/A | N/A | N/A | COMP 01105 | Needs Test | Needs Te.. | FR | 101 |